\documentclass[traditabstract]{aa}

\usepackage{graphicx,amssymb,multirow,gensymb,lscape}
\usepackage{natbib,txfonts}
\newcommand{\vol}{\mbox{cm$^{-3}$}} 
\newcommand{\cden}{\mbox{cm$^{-2}$}} 
\newcommand{\kms}{\mbox{km s$^{-1}$}}
\newcommand{\um}{\mbox{$\mu$m}}
\newcommand{\ammonia}{\mbox{NH$_{3}$}}
\newcommand{\Msun}{\mbox{M$_{\odot}$}}
\newcommand{\cmg}{\mbox{cm$^2$ g$^{-1}$}}
\newcommand{\Av}{\mbox{$A_V$}}
\newcommand{\NHH}{\mbox{N(H$_2$)}}

\newcommand{\vlsr}{\mbox{$v_{lsr}$}}
\newcommand{\HCFN}{\mbox{HC$_5$N}}
\newcommand{\sigt}{\mbox{$\sigma_{T}$}}
\newcommand{\signt}{\mbox{$\sigma_{NT}$}}
\newcommand{\ccs}{\mbox{CCS ($2_1 - 1_0$)}}

\begin{document}

\title{Herschel Observations of a Potential Core-Forming Clump: Perseus B1-E}
\author{S. I. Sadavoy\inst{1,2} \and 
	J. Di Francesco\inst{2,1} \and
	Ph. Andr\'{e}\inst{3} \and
	S. Pezzuto\inst{4} \and
	J.-P. Bernard\inst{5,6} \and
	S. Bontemps\inst{3} \and
	E. Bressert\inst{7,8} \and
	S. Chitsazzadeh\inst{1,2} \and
	C. Fallscheer\inst{1,2} \and
	M. Hennemann\inst{3} \and
	T. Hill\inst{3} \and
	P. Martin\inst{9} \and
	F. Motte\inst{3} \and
	Q. Nguy$\tilde{\hat{\rm e}}$n Lu{\hskip-0.65mm\small'{}\hskip-0.5mm}o{\hskip-0.65mm\small'{}\hskip-0.5mm}ng\inst{3} \and
	N. Peretto\inst{3} \and
	M. Reid\inst{9} \and
	N. Schneider\inst{3} \and
	L. Testi\inst{7,10} \and
	G. J. White\inst{11,12} \and
	C. Wilson\inst{13}
	     }
\institute{Department of Physics and Astronomy, University of Victoria, PO Box 355, STN CSC, Victoria BC Canada, V8W 3P6 \email{ssadavoy@uvic.ca} \and
	National Research Council Canada, Herzberg Institute of Astrophysics, 5071 West Saanich Road, Victoria BC Canada, V9E 2E7 \and
	Laboratoire AIM, CEA/DSM-CNRS-UniversitŽ Paris Diderot, IRFU/Service d'Astrophysique, CEA Saclay, 91191 Gif-sur-Yvette, France \and
	Instituto di astrofisica e planetologia spaziali - INAF, Via del Fosso del Cavaliere, 100, I-00133 Roma, Italy \and
	CNRS, IRAP, 9 Av. colonel Roche, BP 44346, 31028 Toulouse Cedex 4, France \and
	Universit\'{e} de Toulouse, UPS-OMP, IRAP, 31028 Toulouse Cedex 4, France \and
	ESO, Karl Schwarzschild-Strasse 2, D87548 Garching bei Munchen, Germany \and
	School of Physics, University of Exeter, Stocker Road, Exeter EX4 4QL, UK \and
	Canadian Institute for Theoretical Astrophysics, University of Toronto, Toronto, ON, M5S 3H8, Canada \and
	INAF-Osservatorio Astrofisico di Arcetri, Largo E. Fermi 5, 50125 Firenze, Italy \and
	The Rutherford Appleton Laboratory, Chilton, Didcot OX11 0NL, UK \and
	The Open University, Department of Physics and Astronomy, Milton Keynes MK7 6AA, UK \and
	Department of Physics and Astronomy, McMaster University, Hamilton, ON, L8S 4M1, Canada
	     }

\date{Received August 23, 2011; Accepted November 28, 2011}

\abstract
{We present continuum observations of the Perseus B1-E region from the \emph{Herschel} Gould Belt Survey. These \emph{Herschel} data reveal a loose grouping of substructures at $160 - 500$ \um\ not seen in previous submillimetre observations. We measure temperature and column density from these data and select the nine densest and coolest substructures for follow-up spectral line observations with the Green Bank Telescope. We find that the B1-E clump has a mass of $\sim$ 100 \Msun\ and appears to be gravitationally bound. Furthermore, of the nine substructures examined here, one substructure (B1-E2) appears to be itself bound. The substructures are typically less than a Jeans length from their nearest neighbour and thus, may interact on a timescale of $\sim$ 1 Myr. We propose that B1-E may be forming a first generation of dense cores, which could provide important constraints on the initial conditions of prestellar core formation. Our results suggest that B1-E may be influenced by a strong, localized magnetic field, but further observations are still required.}


\authorrunning{Sadavoy et al.}
\titlerunning{Herschel Observations of Perseus B1-E: A First Look at Core Formation}
\maketitle

\section{Introduction\label{Intro}}

Molecular clouds are highly structured regions of dust and gas. They contain dense, small-scale, star-forming ``cores'' ($\lesssim 0.1$ pc) that are usually clustered into larger-scale clumps and organized along filaments \citep{Williams00, Ward-T07, difran07, Andre10}. The cause of this hierarchical structure, where the large-scale clouds ($\sim 10$ pc) with moderate densities ($\sim 10^2$ \vol) produce filaments, clumps, and dense cores at higher densities ($\gtrsim 10^4$ \vol), is not well understood \citep{BerginTafalla07}. 

Molecular clouds form dense cores when diffuse gas is compressed. Current theories for core formation primarily focus on two ideas: that cores form via (1) turbulent compression of diffuse gas \citep[e.g.,][]{Larson81, MacLowKlessen04, Dib09}, or (2) the motion of neutral material between magnetic field lines or ambipolar diffusion \citep[e.g.,][]{MestelSpizter56, Mouschovias76, Kunz09}. Both mechanisms and gravity likely play important roles in star formation, but it is unclear whether one mechanism would dominate in all situations (i.e., clustered or isolated star formation and low-mass or high-mass star formation). Furthermore, theoretical studies also suggest that additional processes, such as radiation feedback \citep[i.e.,][]{Krumholz09} or large-scale shocks associated with converging flows \citep[i.e.,][]{Heitsch11}, can influence molecular cloud structure and thus, core formation. Additionally, recent \emph{Herschel} studies have shown that filaments are very prominent in star forming regions and likely have an important role in the formation of dense substructures in molecular clouds \citep[e.g.,][]{Andre10, Arzoumanian11}.

Surveys of core populations in molecular clouds indicate that cores are confined to regions of high column density. \citet{Johnstone04} and \citet{Kirk06} found a relationship between core occurrence and extinction for the Ophiuchus and Perseus clouds, respectively, suggesting that there is a core formation threshold at $\Av \gtrsim 5$. Similarly, \citet{Lada09} and \citet{Andre10} each compared two clouds with different degrees of star formation activity and each found that the more active cloud was composed of higher column density material (by a factor of $\sim 10$) than the quiescent cloud. These studies emphasize that cores require a minimum column density (extinction) of material to condense from the bulk cloud \citep[see also][]{Heiderman10}.

The precursors to cores are difficult to identify. Dense cores are often influenced by processes such as nearby outflows and radiation feedback from a previous epoch of nearby star formation, and observations of them cannot be used to constrain the dynamic properties of the initial core-forming material \citep{Curtis11}. Without knowing the \emph{initial} conditions and processes that cause cores to form from diffuse gas, we cannot accurately model their formation or evolution. Thus, identifying and analyzing a core forming region without earlier episodes of star formation would be exceedingly useful to constrain how molecular clouds form star-forming substructures. 

In this paper, we use observations from the \emph{Herschel} Gould Belt Survey to explore a $\sim$ 0.1 deg$^2$ clump roughly 0.7$\degree$ east of the B1 clump \citep{Bachiller86} within the Perseus molecular cloud. This region, hereafter called B1-E, has high extinction ($A_V > 5$) similar to the core formation threshold. Despite this high extinction, previous submillimetre and infrared continuum observations suggest that B1-E contains neither dense cores nor young stellar objects \citep[e.g.,][]{Enoch06, Kirk06, Jorgensen07, Evans09}. In contrast, our \emph{Herschel} observations show substructure that was not detected by these other far-infrared and submillimetre continuum studies. Furthermore, we use recent Green Bank Telescope (GBT) observations to quantify in part the kinematic motions of the densest substructures seen in the \emph{Herschel} data. 

We propose that B1-E is forming a first generation of dense cores in a pristine environment. In Section \ref{obs}, we describe our \emph{Herschel} and GBT observations. In Section \ref{results} we describe the properties we derive from our data. In Section \ref{discussion} we discuss the implication of our results and compare our observations to theoretical models. Finally, in Section \ref{conc} we summarize the paper.


\section{Data}\label{obs}

\subsection{\emph{Herschel} Observations}
The western half of Perseus, including B1-E, was observed by the Photodetector Array Camera and Spectrometer \citep[PACS;][]{Poglitsch10} and the Spectral and Photometric Imaging Receiver \citep[SPIRE;][]{Griffin10} as part of the \emph{Herschel} Gould Belt Survey \citep{AndreSaraceno05, Andre10}. The \emph{Herschel} observations were taken in ``parallel mode'', resulting in simultaneous coverage at 5 wavelength bands, with the 70 \um\ and 160 \um\ PACS channels and the 250 \um, 350 \um, and 500 \um\ SPIRE channels, over roughly 6 square degrees. The 70 \um\ observations of B1-E, however, are less sensitive due to the fast scan rate (60\arcsec/s) and low emission from cold material (see Section \ref{seds}). Thus, we will not include the 70 \um\ observations in our discussion. For a full explanation of the observations, see Pezzuto et al. (2012, in preparation). 

\begin{figure*}
\centering
\includegraphics[scale=0.65]{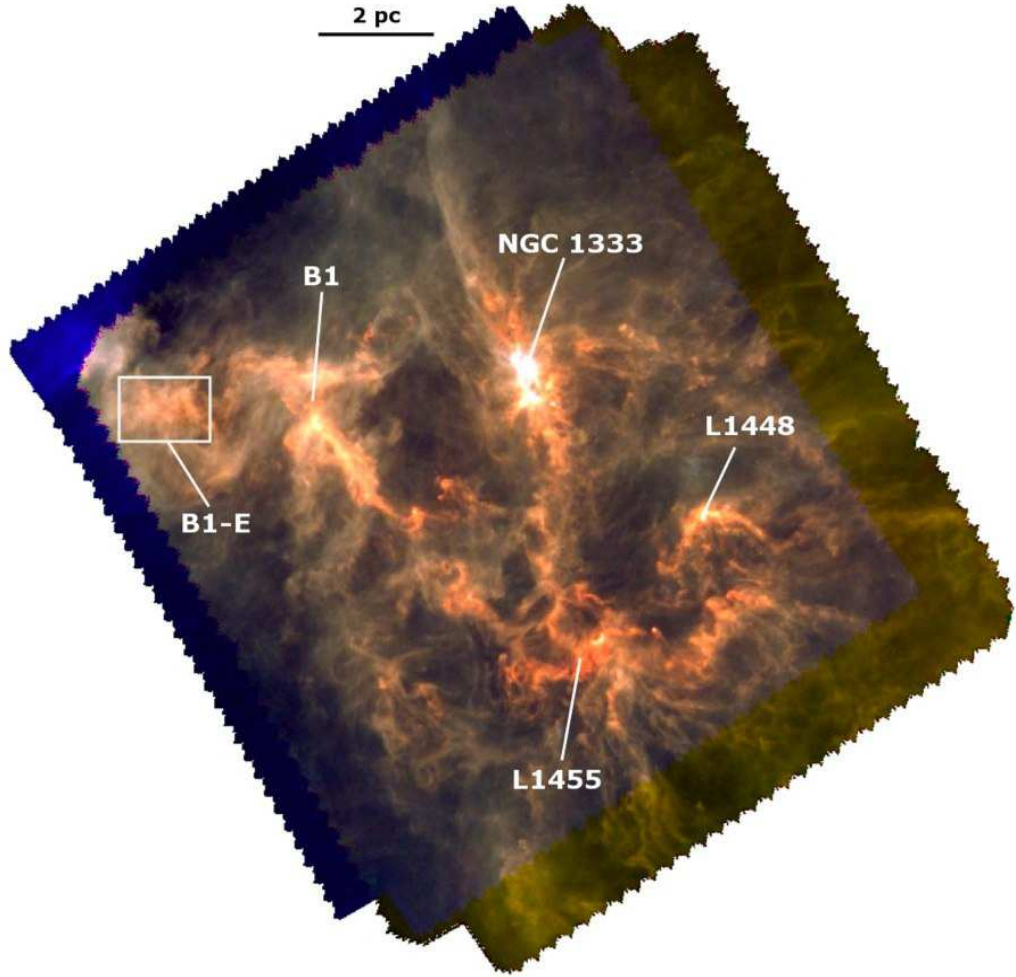}
\caption{Three-colour image of the western half of the Perseus molecular cloud. Colour mosaic was generated using \emph{Herschel} 160 \um, 250 \um, and 350 \um\ observations. The white box denotes our boundary for B1-E for subsequent figures. The other prominent clumps in Perseus are also labeled. \label{west}}
\end{figure*}

The PACS and SPIRE raw data were reduced using HIPE version 5.0 and reduction scripts written by M. Sauvage (PACS) and P. Panuzzo (SPIRE) that were modified from the standard pipeline. For SPIRE reduction, we used updated calibration information (version 4). The final maps were created using the \emph{scanamorphos} routine developed by Roussel (2011, submitted to A\&A). For consistency with other published maps, we set the pixel scale\footnote{By default, \emph{scanamorphos} sets a much smaller pixel scale.} to the default size from the HIPE mapmaking tools, which results in pixels of sizes 6.4$\arcsec$, 6.0$\arcsec$, 10.0$\arcsec$, and 14.0$\arcsec$ for 160 \um, 250 \um, 350 \um, and 500 \um, respectively. We also adopt beam sizes of 13.4$\arcsec$, 18.1$\arcsec$, 25.2$\arcsec$, and 36.6$\arcsec$ for the 160 \um, 250 \um, 350 \um, and 500 \um\ bands respectively\footnote{Due to the fast scan rate, the 160 \um\ beam is slightly elongated along the scan direction. Thus, we adopt the geometric mean (i.e., $r = \sqrt{ab}$) as the beam size, where the elongated beam dimensions are from \citet{Poglitsch10}.} \citep[see][]{Griffin10, Poglitsch10}. Assuming a distance to Perseus of 235 pc \citep{Hirota08}, \emph{Herschel} can detect structure on scales of $\sim 0.04$ pc at 500 \um. Figure \ref{west} shows a three-colour image of Western Perseus with labels for B1-E and other prominent subregions.

\begin{figure*}
\includegraphics[scale=0.5]{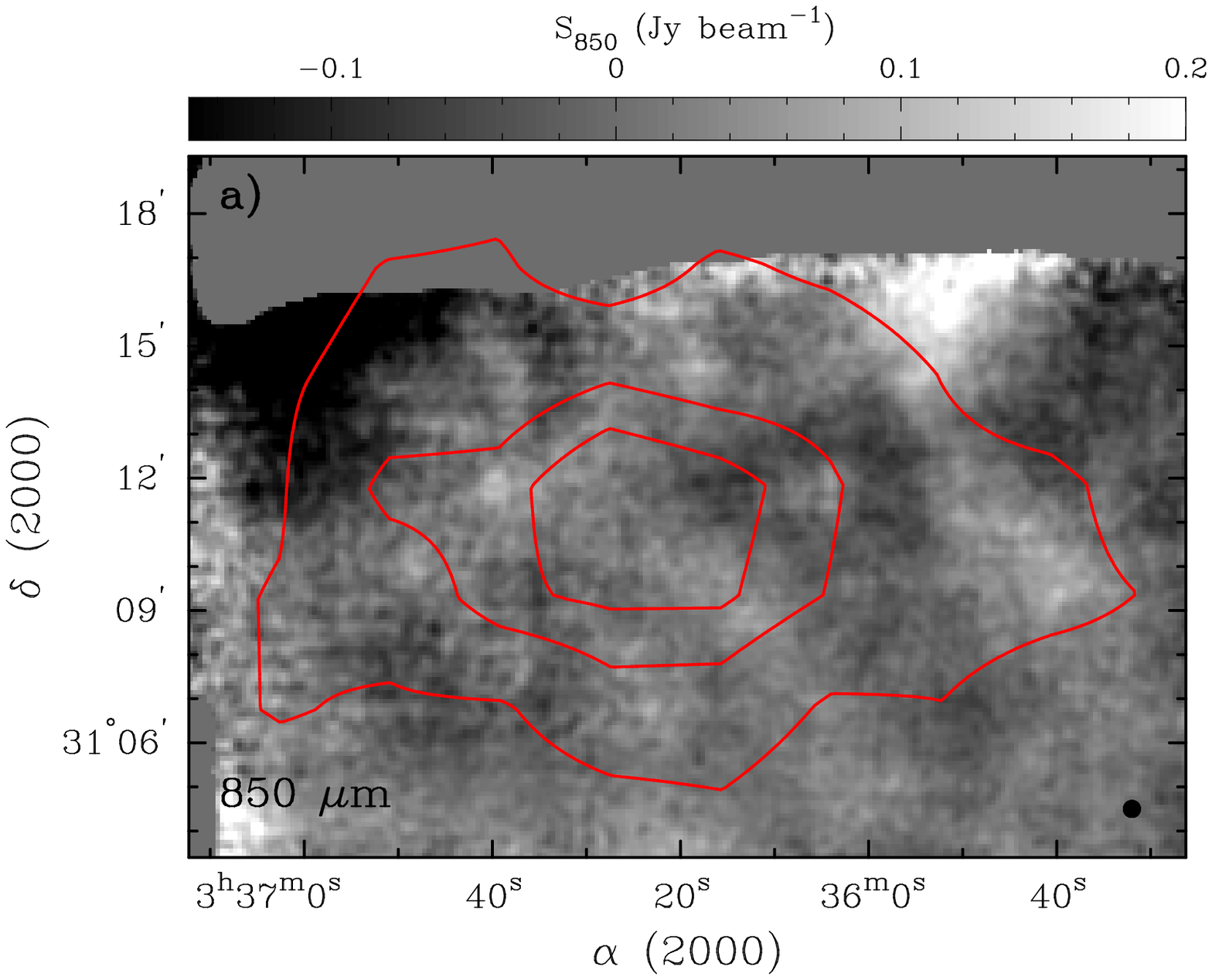} \hspace{2pt} \includegraphics[scale=0.5]{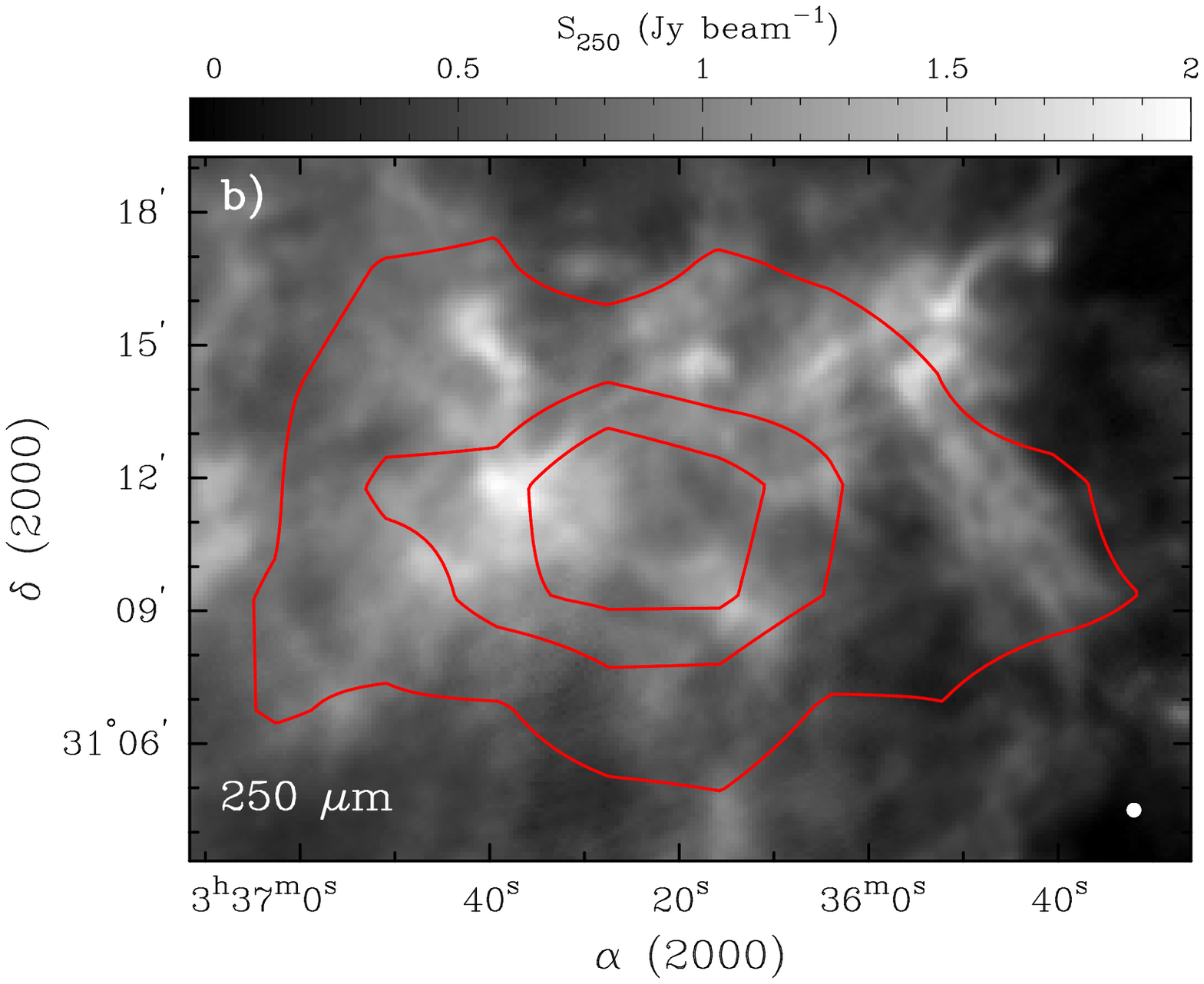}
\caption{Observations of Perseus B1-E from (a) SCUBA 850 \um\ and (b) SPIRE 250 \um\ maps. The SCUBA data came from the Extended SCUBA Legacy Catalogue \citep[see][]{difran08}. Contours represent extinction levels of $A_V = 5$, 7, 8 magnitudes from the COMPLETE extinction map \citep{Ridge06} and the filled circles show the beam sizes of 23$\arcsec$ for smoothed SCUBA data at 850 \um\ and 18$\arcsec$ for SPIRE data at 250 \um.\label{b1e850_250}}
\end{figure*}

Unlike previous ground-based submillimetre instruments, \emph{Herschel} can detect large-scale diffuse emission \citep[e.g.,][]{Schneider10, Miville-D10, Arzoumanian11}. Additionally, \emph{Herschel} has excellent sensitivity to low-level flux. For example, Figure \ref{b1e850_250} compares SCUBA 850 \um\ observations of B1-E \citep[smoothed to $23\arcsec$ resolution;][]{difran08} with the new SPIRE 250 \um\ observations (at $18\arcsec$ resolution). The SPIRE 250 \um\ data show prominent substructures not identified in the SCUBA 850 \um\ data, though several faint features at 850 \um\ appear to agree with the brighter structures in the 250 \um\ map. With the higher sensitivity of \emph{Herschel}, we are able to identify clearly structures in B1-E that were too faint, i.e., $< 3\ \sigma$, to be robust detections with SCUBA.

\subsection{GBT Observations}

We selected the nine highest peaks in our \emph{Herschel}-derived H$_2$ column density map of B1-E (see Section \ref{CDprofiles}) for complementary follow-up observations with the new K-band Focal Plane Array (KFPA) receiver on the GBT\footnote{The GBT is an 100 m telescope operated by the National Radio Astronomy Observatory.}. Our targets, hereafter named B1-E1 to B1-E9 according to decreasing peak column density, were observed on 03 March 2011 with single-pointings. We used the KFPA receiver with one beam and four spectral windows to observe each target simultaneously in \ammonia\ (1,1), \ammonia\ (2,2), \ccs, and \HCFN\ (9-8) line emission at 23.6945 GHz, 23.7226 GHz, 22.3440 GHz, and 23.9639 GHz, respectively. Similar to \citet{Rosolowsky08}, we made frequency-switching observations with 9-level sampling and 12.5 MHz bandwidth over 4096 spectral channels in each window to achieve a high velocity resolution of $v_{ch} \approx 0.0386$ \kms\ at 23.69 GHz. B1-E1 to B1-E8 were observed for $\sim 1120$ seconds, each. B1-E9 was observed for $\sim 840$ seconds.

The GBT data were reduced using standard procedures in GBTIDL\footnote{GBTIDL is a special IDL package specific to the GBT.} for frequency-switched data.  In brief, individual scans from each spectral window were filtered for spikes or baseline wiggles, folded, and then averaged.  Baselines were obtained for the averaged spectra by fitting 5th-order polynomials to line-free channel ranges at the low- and high-frequency edges of each band, and were then subtracted.  To improve the detection levels, the data were smoothed with a boxcar kernel equal to two channels in width.  The reduced data were exported from GBTIDL into standard FITS files using routines of AIPS$^{++}$ developed by G. Langston.  Further analysis was conducted in MIRIAD\footnote{Multichannel Image Reconstruction, Interactive Analysis, and Display (MIRIAD) software is developed by the Berkeley Illinois Maryland Array (BIMA) group. See \citet{Sault95} for more details.} and IDL (Interactive Data Language). The 1 $\sigma$ noise levels were typically $\sim 0.01 - 0.02$ K. At 23.69 GHz, the GBT beam is $\sim 33\arcsec$ and the beam efficiency is $\eta_b = 0.825$.

Table \ref{pointings} gives the positions, peak H$_2$ column density, and the detected line emission of these targets. \ammonia\ (1,1) line emission was detected ($> 3\ \sigma$) towards all nine targets and \ccs\ line emission towards several. The \ammonia\ (2,2) and \HCFN\ (9-8) lines were only detected toward B1-E2. 

\begin{table*}
\caption{GBT Target Information}
\label{pointings}
\centering
\begin{tabular}{lcccp{8.2cm}}
\hline\hline
Source & $\alpha$ (J2000) & $\delta$ (J2000) & Peak \NHH\tablefootmark{a} & Detections\tablefootmark{b}\\
    & (h:m:s)  & ($\degree\ \arcmin\ \arcsec$) & ($10^{21}$ \cden) & \\
\hline
B1-E1 & 3:35:55.0 & 31:14:16 & 18.5 & \ammonia\ (1,1), \ccs \\
B1-E2 & 3:36:04.4 & 31:11:47 & 16.7 & \ammonia\ (1,1), \ammonia\ (2,2), \ccs, \HCFN\ (9-8)\\
B1-E3 & 3:35:52.1 & 31:15:39 & 15.1 & \ammonia\ (1,1), \ccs \\
B1-E4 & 3:35:51.0 & 31:12:34 & 14.9 & \ammonia\ (1,1)\\
B1-E5 & 3:36:37.3 & 31:11:41 & 14.5 & \ammonia\ (1,1)\\
B1-E6 & 3:36:41.0 & 31:15:05 & 13.7 & \ammonia\ (1,1)\\
B1-E7 & 3:36:39.0 & 31:14:27 & 13.6 & \ammonia\ (1,1), \ccs \\
B1-E8 & 3:36:05.0 & 31:14:28 & 13.1 & \ammonia\ (1,1), \ccs \\
B1-E9 & 3:36:18.5 & 31:14:31 & 12.9 & \ammonia\ (1,1)\\
\hline
\end{tabular}
\tablefoot{
\tablefoottext{a}{Peak \emph{Herschel}-derived $H_2$ column density towards the sources. See Section \ref{results} for more details.} 
\tablefoottext{b}{Line emission detected towards each source.}
}
\end{table*}

\section{Results}\label{results}

\subsection{SED Fitting to \emph{Herschel} Data}\label{seds}
We corrected the arbitrary zero-point flux offset in each \emph{Herschel} band using the method proposed in \citet{Bernard10} that is based on a comparison with the Planck HFI \citep[DR2 version, see][]{Planck_HFI} and IRAS data. In addition, each map was convolved to the resolution of the 500 \um\ map ($36.6\arcsec$) and regridded to $14\arcsec$ pixels. The map intensities of the $160 - 500$ \um\ bands were then fit by the modified black body function,
\begin{equation}
I_{\nu} = \kappa_{\nu}B_{\nu}(T)\Sigma \label{modBB}
\end{equation}
where $\kappa_{\nu}$ is the dust opacity, $B_{\nu}$ is the black body function, $T$ is the dust temperature, and $\Sigma$ is the gas mass column density. Note that $\Sigma = \mu m_H\NHH$, where $\mu$ is the mean molecular weight, $m_H$ is the hydrogen mass, and \NHH\ is the gas column density. For consistency with other papers in the Gould Belt Survey \citep[e.g.,][]{Andre10},
\begin{equation}
\kappa_{\nu} = 0.1(\nu/1000\ \mbox{GHz})^{\beta}\ \cmg, \label{kappaEq}
\end{equation}
where $\beta$ is the dust emissivity index. The SED fits were made using the IDL program \emph{mpfitfun} by C. B. Markwardt. In brief, \emph{mpfitfun} performs a least-squares comparison between the data and a model function by adjusting the desired parameters until a best fit is achieved.

\indent Fitting $\beta$ requires a plethora of data, particularly along the Rayleigh-Jeans tail (e.g., $\lambda > 300$ \um\ at 10 K), to remove degeneracies in the model fits \citep{Doty94, Shetty09}. Since we have only two photometric bands (350 \um\ and 500 \um) along the Rayleigh-Jeans tail, we assume $\beta = 2$, consistent with the adopted $\beta$ in other \emph{Herschel} first-look studies \citep[e.g.,][]{Andre10, Arzoumanian11}. There is some evidence for $\beta \approx 2$ in recent Planck studies \citep[see][and references therein]{Planck_beta1, Planck_beta2}. Adopting $\beta = 2$ here provides good fits to our data (see below). 

Figure \ref{temp_CD} shows the dust temperatures and column densities across B1-E resulting from the modified black body fits to each pixel for the $160-500$ \um\ bands, assuming flux uncertainties of 15\% based on calibration uncertainties \citep[see][]{Griffin10, Poglitsch10} and $\beta = 2$. Both temperature and column density in B1-E are highly structured (see Figures \ref{temp_CD}a and \ref{temp_CD}b), where regions of higher column density are also associated with slightly cooler temperatures and regions of lower column density have slightly warmer temperatures. Most of the cooler, high column density material ($\gtrsim 6 \times 10^{21}$ \cden) is clumped into a central ring-like structure (i.e., slight depression is seen towards the very centre).

\begin{figure*}
\includegraphics[scale=0.625,angle=-90]{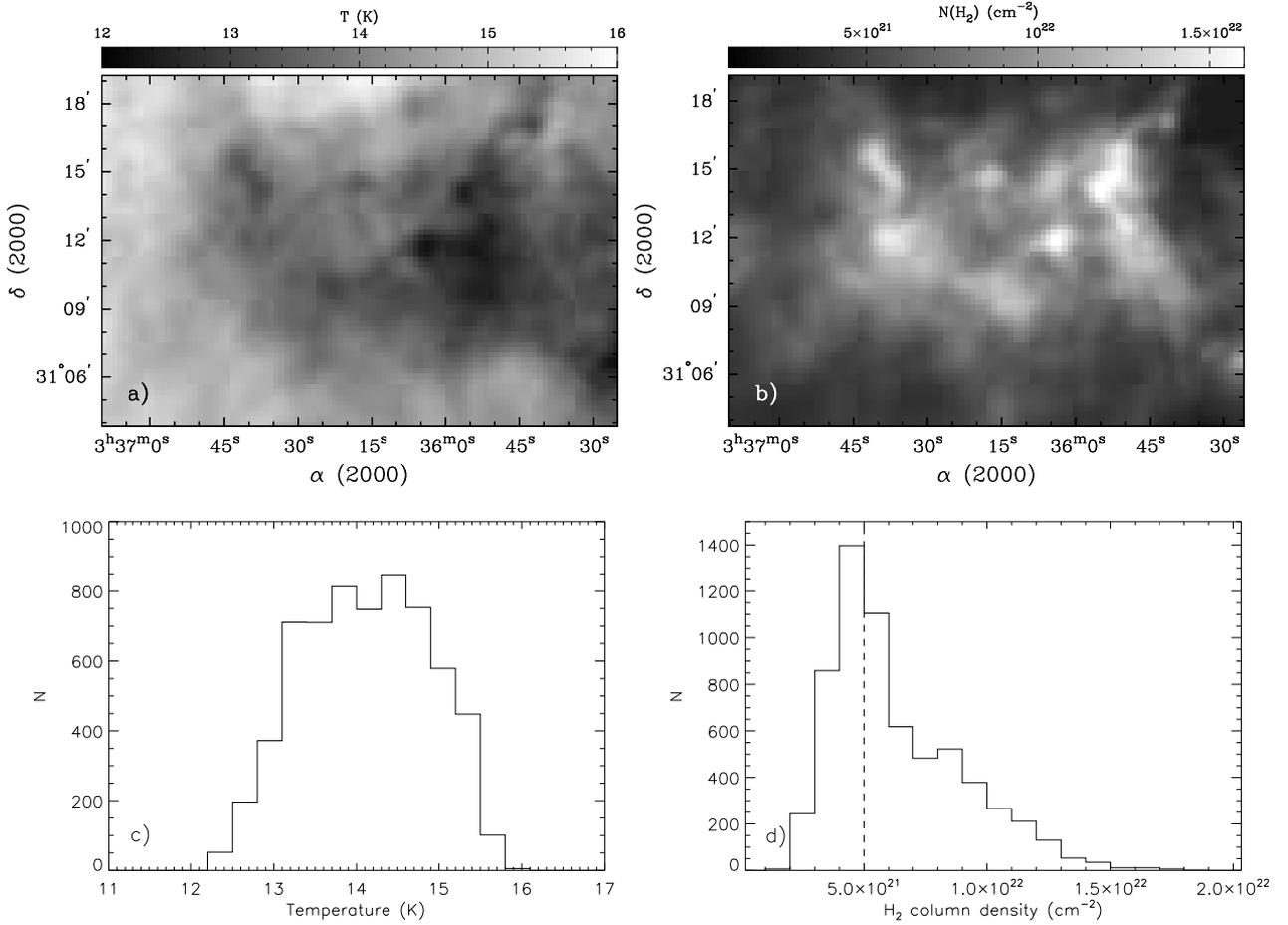}
\caption{SED-fitting results for B1-E. \emph{Top panels} shows the temperature (a) and H$_2$ column density (b) maps across B1-E as measured from SED-fitting to 160 - 500 \um\ data, according to Equations \ref{modBB} and \ref{kappaEq} and assuming $\beta = 2$. \emph{Bottom panels} show the histograms of the above maps. The temperature histogram (c) uses a bin size of 0.3 K and the H$_2$ column density histogram (d) uses a bin size of  $1 \times 10^{21}$ \cden. For comparison, the dashed line in panel (d) indicates the observed column density threshold (from extinction) for core formation from \citet{Kirk06}.\label{temp_CD}}
\end{figure*}

Figures \ref{temp_CD}c and \ref{temp_CD}d show the number histograms of temperature and H$_2$ column density, respectively. These distributions are non-Gaussian. The sample mean and standard deviation about the mean for the temperature and column density are $14.1 \mbox{ K} \pm 0.8$ K and $(6.3 \pm 2.7) \times 10^{21}$ \cden, respectively, where the standard deviation is computed from,
\begin{equation}
SE = \sqrt{\frac{1}{N-1}\sum\left(x_i - \bar{x}\right)^2},
\end{equation}
where $\bar{x}$ is the population mean for the sample of size $N$.

These mean values agree well with previously estimated quantities for this region. For example, \citet{Schnee05} measured a mean dust temperature of $\sim 14$ K for the B1-E region using the ratio of IRAS 60 \um\ and 100 \um\ flux densities to estimate dust temperature. Additionally, extinction data from the COMPLETE survey (see contours in Figures \ref{b1e850_250} and \ref{temp_CD}) suggest a mean column density of $(5.3 \pm 1.5) \times 10^{21}$ \cden, assuming $\NHH/\Av = 10^{21}\ \cden$ mag$^{-1}$. Using a higher resolution extinction map from S. Bontemps, we find a mean column density of $(6.3 \pm 3.6) \times 10^{21}$ \cden. These two column densities agree very well with our measured value of $(6.3 \pm 2.7) \times 10^{21}$ \cden\ and they are also similar to the threshold column density for dense core formation from the extinction analysis in Perseus by \citet{Kirk06}, $5 \times 10^{21}$ \cden. For comparison, Figure \ref{temp_CD}d includes the Kirk et al. column density threshold as a dashed line. Although B1-E does not appear filamentary, there is ample material from which dense cores may form. With higher resolution data, \citet{Andre11} found a threshold column density of $7 \times 10^{21}$ \cden\ for dense structures to form in Aquila via thermal instabilities along a filament at 10 K.

\subsection{Column Density Profiles} \label{CDprofiles}

One of the more prominent core models is the Bonnor-Ebert sphere \citep{Bonnor56, Ebert55}, which represents the density profile of a sphere in hydrostatic equilibrium under the influence of an external pressure. The Bonnor-Ebert sphere has a flat inner density distribution and a power-law density downtrend at larger radii. Many prestellar cores, i.e., dense cores that are gravitationally bound but do not show evidence of a central luminous protostar, have shown Bonnor-Ebert-like profiles \citep[e.g.,][]{Ward-T99, Alves01}.

With the excellent spatial resolution of \emph{Herschel}, we can measure the column density profiles for individual B1-E substructures. First, we identified the \emph{locations} of peak column density using the 2D Clumpfind algorithm \citep{Clumpfind}. Briefly, Clumpfind identifies intensity peaks and then uses closed contours at lower intensity levels to assign boundaries. We will discuss the nine highest column density substructures identified by Clumpfind (B1-E1 to B1-E9) for this paper. Second, we measured the azimuthally-averaged column density profile of our nine sources using the \emph{ellint} task in MIRIAD. For simplicity, we used circular annuli of 10$\arcsec$ width for $r > 7\arcsec$. For the central radii ($r < 7\arcsec$), we assume the peak column density. We caution that several of our sources appear elliptical and our circular approximation is meant to provide a broad, first-look analysis.

\begin{figure*}
\includegraphics[scale=0.875]{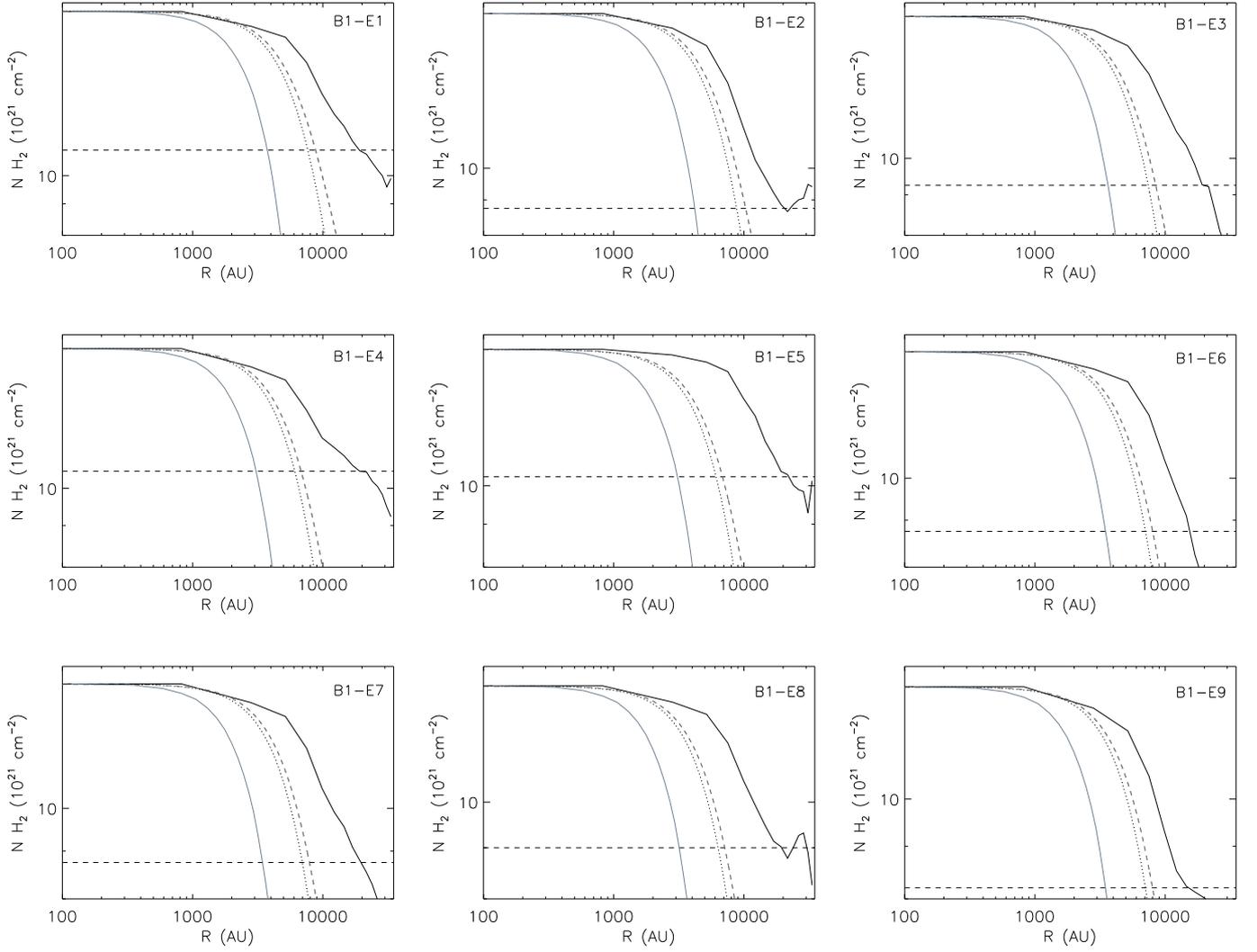}
\caption{The azimuthally-averaged column density profiles for the nine sources in B1-E, assuming a distance of 235 pc. The profiles were measured using circular annuli with a thickness of 10$\arcsec$ for $r > 7\arcsec$. Note that the central area is defined by a circle of radius 7$\arcsec$ but plotted at 3.5$\arcsec$. The dashed line indicates our estimate for the non-source line-of-sight (LOS) material. For comparison, the dotted curve and dashed curve illustrate a Bonnor-Ebert column density profile and a ``generic'' $[1 + (r/a)^2]^{-0.5}$ column density profile following \citet{DappBasu09}. For both analytic curves, we assume a temperature of 10 K and a central density of $n = 10^6$ \vol\ and convolved the profiles with a 36.6$\arcsec$ beam. The grey solid curve shows the beam profile. The analytical profiles and the beam profile are scaled to the peak column density. Note that both axes are logarithmic. \label{example_profiles}}
\end{figure*}

Figure \ref{example_profiles} shows the column density profiles, where the column density values are plotted from the centre of each annulus. For comparison, Figure \ref{example_profiles} also shows the beam profile (solid grey curve),  a ``generic'' column density profile convolved with the beam (dashed grey curve) and an analytical Bonnor-Ebert profile convolved with the beam (dotted curve). For both analytic profiles, we follow the approximation from \citet{DappBasu09} and assume a temperature of $T = 10$ K, a central density of $n = 10^6$ \vol\ and a constant of proportionality of $k = 0.54$, where $k \approx 0.4$ for a singular isothermal sphere and $k \approx 1$ for a collapsing cloud. Our observed column density profiles are much wider than the models, likely due to contaminating material in the foreground or far-background along the line of sight, hereafter called LOS material. Towards the centre of each source, the analytic models and observed profiles are more similar, since the source column density dominates over the LOS level, whereas in the power-law roll-off, the LOS material is likely more significant and the profiles deviate from the analytic models. Unfortunately, the column density of such extended material is difficult to distinguish from the source column density. To estimate its contribution towards each source, we used the subsequent column density at the location where the power-law slope in the column density profile flattened or began to increase. Thus, the LOS material level ranges from $\sim 8 \times 10^{21}$ \cden\ for B1-E9 to $\sim 11 \times 10^{21}$ \cden\ for B1-E1 or roughly 60\%\ of the peak column density. These LOS column densities are very conservative and could overestimate their actual contributions by as much as a factor of 2. Indeed, these values are larger than the mean B1-E column density of $6.3 \times 10^{21}$ \cden.

\begin{figure*}
\centering
\includegraphics[scale=0.875]{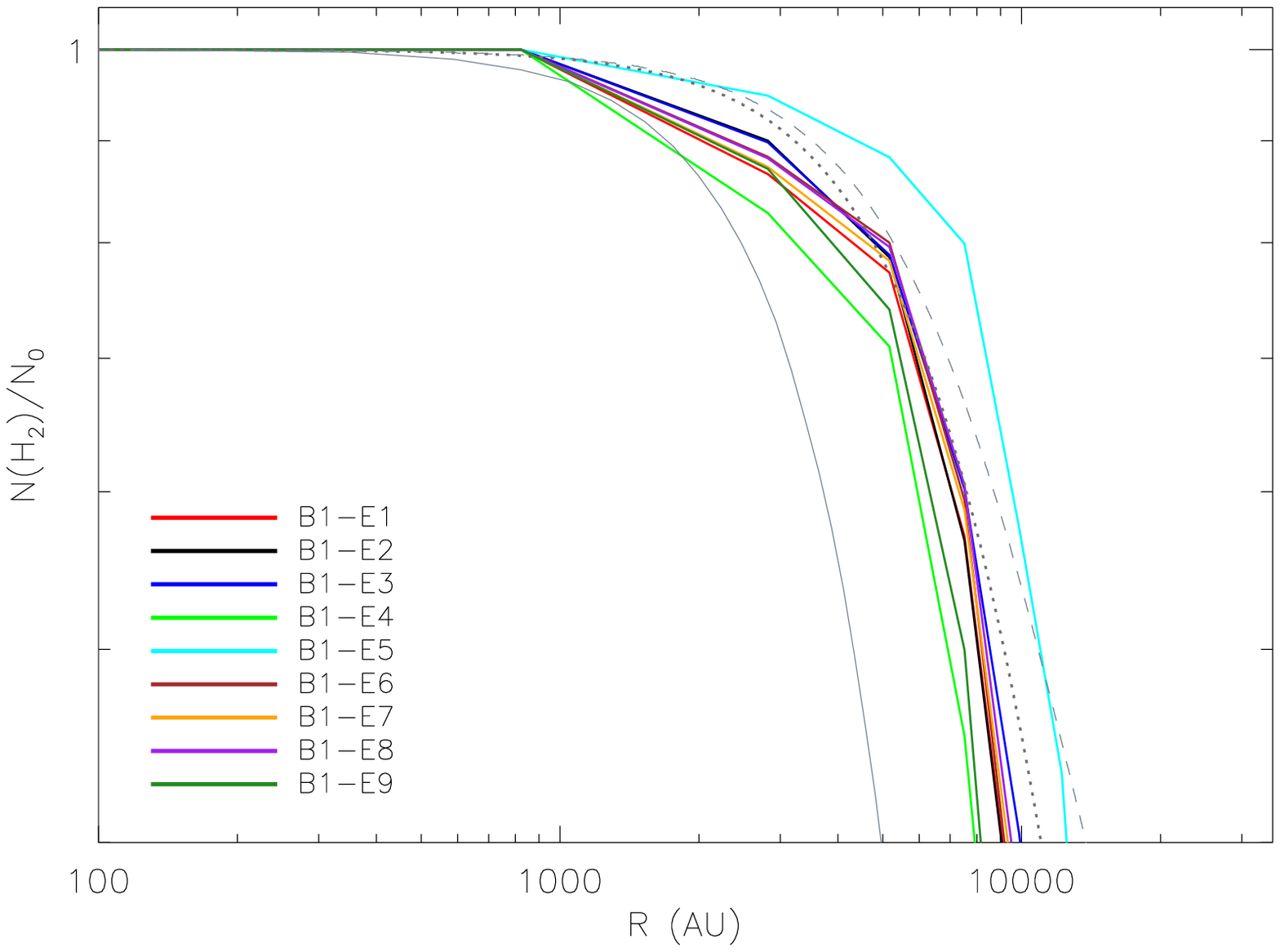}
\caption{Normalized column density profiles for sources in B1-E, assuming a distance of 235 pc. Column density was determined from azimuthally-averaged annuli (see Figure \ref{example_profiles}). Furthermore, the LOS material was estimated for each source and subtracted from the azimuthal average. The grey solid, dotted, and dashed curves are the same as for Figure \ref{example_profiles}.\label{profiles}}
\end{figure*}

Figure \ref{profiles} shows a comparison of our nine column density profiles after correcting each profile for LOS material and normalizing to the respective peak column density, and includes the Bonnor-Ebert profile, generic profile, and beam profile from Figure \ref{example_profiles}. All nine profiles follow a shape similar to those of the analytic models, with a flat centre and steep falloff towards larger radii. Additionally, all profiles but B1-E5 generally appear more centrally concentrated than the models, implying that B1-E5 is the least compact. The column density profiles do show some differences with the models, however, such as steeper fall-offs. 

\subsection{Substructures}\label{substructures}

We estimated source sizes and masses based on Gaussian fits to the LOS-subtracted column density profiles.  Source size was defined by the FWHM of the Gaussian (i.e., $R = \mbox{FWHM}/2$). Since B1-E6 and B1-E7 have a small projected separation of $\sim 50\arcsec$, we truncated both to $31\arcsec$ radii\footnote{Deconvolved radii are $\approx 25\arcsec$.} to ensure that we mostly measured column densities unique to each. Thus, our analyses of B1-E6 and B1-E7 are biased toward the denser, central regions unlike those of the other sources. Subtracting out the LOS material limits the influence from nearby extended material and generally fits only the denser, embedded object. If we include the LOS material, the estimated source sizes generally increase by $\lesssim 20\%$. Isolated substructures, like B1-E2 or B1-E8, have sizes that vary by $\sim 5\%$ and substructures embedded in extended profiles like, B1-E1 or B1-E4, have sizes that vary by $24\%$ and $37\%$, respectively.

Table \ref{properties} lists the adopted properties for our sources, including their deconvolved radii, estimated (upper and lower) masses, and average densities, assuming they are perfect spheres and dividing the (upper and lower) source masses with their spherical volumes, $\frac{4}{3}\pi R_d^3$, where $R_d$ is the deconvolved radius. We consider two mass and density limits: the lower limits subtract out the LOS material and the upper limits include the LOS material. Note that, by our definition of source size, we are measuring the inner regions of each profile where core precursors are most likely to arise. Since we are using the Gaussian FWHM to estimate the source size, the  true source sizes, including any diffuse envelopes, may be larger by a factor of 2. Source mass was estimated by summing over the column density and assuming a mean molecular weight of $\mu = 2.33$.  These results provide a reasonable first look at the \emph{relative} sizes and masses of these sources. 

\begin{table}
\caption{Column Density Determined Properties}
\label{properties}
\centering
\begin{tabular}{lccc}
\hline\hline
Source & R$_{d}$\tablefootmark{a}   & M\tablefootmark{b} & $n$\tablefootmark{c}\\
              &          (AU) 	& (\Msun) 			   & ($10^4$ \vol) \\
\hline
B1-E1 & 7.1 $\times 10^3$ & $0.6 - 1.6$ & $5.9 - 17$ \\
B1-E2 & 6.9 $\times 10^3$ & $0.6 - 1.4$ & $6.7 - 15$  \\
B1-E3 & 7.7 $\times 10^3$ & $0.5 - 1.5$ & $4.0 - 12$ \\
B1-E4 & 6.0 $\times 10^3$ & $0.3 - 1.0$ & $4.3 - 18$ \\
B1-E5 & 9.3 $\times 10^3$ & $0.5 - 2.0$ & $2.4 - 9.1 $\\
B1-E6\tablefootmark{d} & 5.9 $\times 10^3$ & $0.3 - 1.0$ & $5.7 - 17$ \\
B1-E7\tablefootmark{d} & 5.9 $\times 10^3$ & $0.3 - 0.9$ & $5.6 - 17$ \\
B1-E8 & 7.2 $\times 10^3$ & $0.3 - 1.2$ & $3.1 - 12$ \\
B1-E9 & 5.4 $\times 10^3$ & $0.3 - 0.8$ & $5.8 - 18$\\
\hline
\end{tabular}
\tablefoot{
\tablefoottext{a}{Object radii were estimated from Gaussian fits to the column density profile ($R = \mbox{FWHM}/2$). Radii were deconvolved with a 36.6$\arcsec$ beam.}
\tablefoottext{b}{Masses are taken from integrating out the column density profiles. The lower limit measurements subtract the LOS material from the column density profiles. The upper limit measurements do not subtract the LOS material from the column density profiles.}
\tablefoottext{c}{Average density determined from the mass and volume.}
\tablefoottext{d}{B1-E6 and B1-E7 are truncated to measure column density \emph{unique} to each source.}
}
\end{table}

\subsection{Line Emission}

\begin{figure}[!h]
\resizebox{\hsize}{!}{\includegraphics[trim=6mm 1.2cm 8mm 6mm,clip=true,angle=-90]{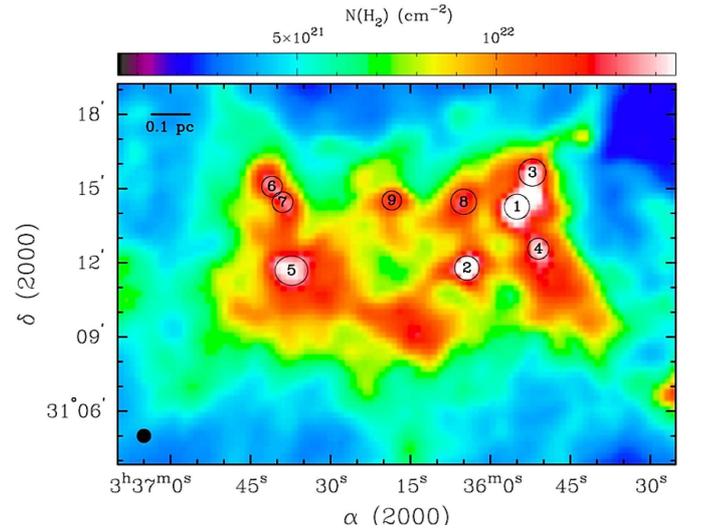}}
\caption{Locations of substructures in B1-E overlaid on our \emph{Herschel}-derived column density map. Numbers indicate the positions of peak column densities and the relative magnitude of the column density peak (see also Table \ref{pointings}). The thin black circles show the deprojected source sizes (see Section \ref{substructures}). The dark filled circle shows the GBT beam ($\sim 33\arcsec$) at 23.69 GHz. \label{targets}}
\end{figure}

\citet{Kirk07} and \citet{Rosolowsky08} each attempted to identify dense gas towards B1-E via high spatial resolution line observations in N$_2$H$^+$ and \ammonia, respectively. They observed several ``blind'' pointings towards B1-E and found no strong detections. Unlike previous continuum studies, our \emph{Herschel} data have identified several cold, dense substructures in B1-E (see Figure \ref{temp_CD}) and these were not directly probed by either Kirk et al. or Rosolowsky et al. Using our \emph{Herschel} results, we selected the nine substructures with the highest column densities for follow-up observations with the GBT in \ammonia\ (1,1), \ammonia\ (2,2), \ccs, and \HCFN\ (9-8) line emission. Figure \ref{targets} shows the locations of our nine targets (see also Table \ref{pointings}). 

\subsubsection{\ammonia}\label{ammoniaSect}
\begin{figure*}
\includegraphics[scale=0.8,angle=-90]{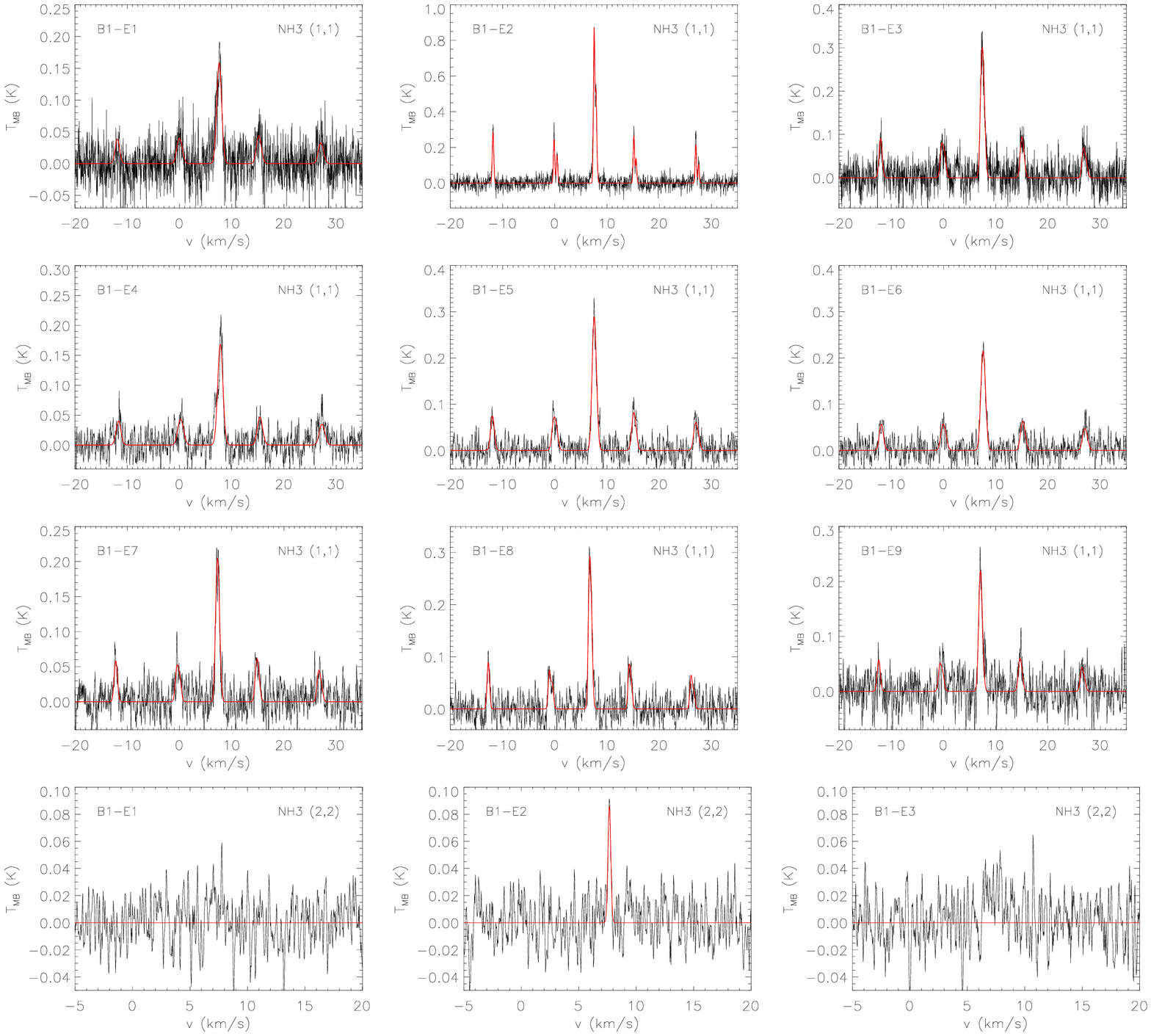}
\caption{Fits to \ammonia\ (1,1) and (2,2) spectra. Note the hyperfine components visible for the \ammonia\ (1,1) spectrum of B1-E2. The \ammonia\ (2,2) spectra were smoothed by a boxcar with a 2-channel width to improve the visual appearance. For comparison, we show two undetected \ammonia\ (2,2) observations.\label{ammoniaSpectra}}
\end{figure*}

We detected \ammonia\ (1,1) emission towards all nine targets. We fit the \ammonia\ (1,1) hyperfine lines with multiple Gaussian components using the following equation;\\

\begin{equation}
\tau(v) = \tau_{1,1} \sum_{i=1}^{18}{\alpha_i \exp{\left[-4\ln{2}\left(\frac{v - v_i - \vlsr}{\Delta{v}}\right)^2\right]}}
\end{equation}
where $\tau_{1,1}$ is the optical depth for the (1,1) transition, $\alpha_i$ is the transition weight, $v_i$ is the velocity of the hyperfine line (for a given rest frequency), \vlsr\ is the velocity of the source with respect to the local standard of rest, and $\Delta{v}$ is the FWHM of the line. We used the hyperfine frequencies and weights from \citet{Kukolich67} and \citet{Rydbeck77} and derived the values for $\tau_{1,1}$, \vlsr, and $\Delta{v}$ from the fits. Figure \ref{ammoniaSpectra} shows example fits to the \ammonia\ spectra. Table \ref{ammonia} lists centroid velocity (\vlsr) and \emph{uncorrected} velocity line width ($\Delta{v}$) from the \ammonia\ (1,1) fits, the kinetic gas temperature from \ammonia\ (1,1) and (2,2) line comparisons, and the resulting thermal and non-thermal velocity dispersions for all nine sources. 

\begin{table*}
\caption{Properties from \ammonia\ (1,1) and \ammonia\ (2,2) Line Spectra}
\label{ammonia}
\centering
\begin{tabular}{lccccc}
\hline\hline
Source &  \vlsr & $\Delta{v}$\tablefootmark{a} & $T_K$\tablefootmark{b} & {\sigt}\tablefootmark{c} & \signt \\
              & (\kms) & (\kms)       & (K) & (\kms) & (\kms)\\
\hline
B1-E1 & 7.61 $\pm$ 0.02 & 0.994  $\pm$ 0.055 & $< 27$ & 0.073 $\pm$ 0.007 & 0.415 $\pm$ 0.025\\
B1-E2 & 7.67 $\pm$ 0.002 & 0.293 $\pm$ 0.005 & 10.27 $\pm$ 0.38 & 0.071  $\pm$ 0.001 & 0.101 $\pm$ 0.003\\
B1-E3 & 7.47 $\pm$ 0.01 & 0.681 $\pm$ 0.030  & $< 16$ & 0.073 $\pm$ 0.007 & 0.280 $\pm$ 0.015\\
B1-E4 & 7.82 $\pm$ 0.01 & 1.080 $\pm$ 0.039 & $< 27$ & 0.073 $\pm$ 0.007 & 0.452 $\pm$ 0.018\\
B1-E5 & 7.57 $\pm$ 0.01 & 0.832 $\pm$ 0.022  & $< 19$ & 0.073 $\pm$ 0.007 & 0.344 $\pm$ 0.011\\
B1-E6 & 7.61 $\pm$ 0.01 & 0.905 $\pm$ 0.032 & $< 23$ & 0.073  $\pm$ 0.007 & 0.376 $\pm$ 0.015\\
B1-E7 & 7.31 $\pm$ 0.01 & 0.708 $\pm$ 0.032  & $< 24$ & 0.073 $\pm$ 0.007 & 0.290 $\pm$ 0.016\\
B1-E8 & 6.78 $\pm$ 0.01 & 0.537 $\pm$ 0.021  & $< 20$ & 0.073 $\pm$ 0.007 & 0.213 $\pm$ 0.012\\
B1-E9 & 7.11 $\pm$ 0.01 & 0.688 $\pm$ 0.038  & $< 32$ & 0.073 $\pm$ 0.007 & 0.281 $\pm$ 0.018\\
\hline
\end{tabular}
\tablefoot{
\tablefoottext{a}{The velocity FWHM from the best-fit models uncorrected for channel resolution.}
\tablefoottext{b}{Kinetic temperature towards each source. For B1-E2, we measured the kinetic temperature from \ammonia\ (1,1) and \ammonia\ (2,2) line emission. For our other sources, we used a 3 $\sigma$ upper limit for \ammonia\ (2,2) to estimate the upper limit of $T_K$.}
\tablefoottext{c}{The thermal line width of B1-E2 was determined using the derived kinetic temperature, $T_K = 10.27 \mbox{ K} \pm 0.38$ K. For the remaining sources, we adopted 11 K, the mean \ammonia-derived kinetic temperature from \citet[][]{Rosolowsky08} with an uncertainty of 2 K from the dispersion of the kinetic temperature in that survey \citep[see][]{Enoch08}.}
}
\end{table*}
\normalsize

For B1-E2, we could derive the kinetic gas temperature using \ammonia\ (1,1) and \ammonia\ (2,2) line emission, whereas we could only determine upper limits for our other sources. We use our \ammonia-derived kinetic temperature of $T_K = 10.27$ K $\pm\ 0.38$ K for B1-E2 and adopt the mean \ammonia-derived kinetic temperature of 11 K from the survey of \citet[][]{Rosolowsky08} for our other sources. As an estimate of the uncertainty, we use the dispersion in the kinetic temperatures from the Rosolowsky et al. data, 2 K, \citep[see][]{Enoch08}. Note that these gas temperatures correspond to the denser, colder inner layers where ammonia is excited, whereas the dust temperatures (see Section \ref{seds}) represent an average of all the material along the line of sight, i.e., including the lower density envelopes. For more information on extracting kinetic temperatures from \ammonia\ line emission, see \citet{Friesen09}.

Table \ref{ammonia} also lists the thermal velocity dispersion, \sigt, and the non-thermal velocity dispersion, \signt. We determined \sigt\ from the kinetic temperature and \signt\ from the line widths via,
\begin{eqnarray}
\sigt\ &=& \sqrt{\frac{k_bT_K}{\mu_{NH_3}m_H}}\\
\signt &=& \sqrt{\sigma_{obs}^2 - \sigt^2},
\end{eqnarray}
where $k_b$ is the Boltzmann constant, $T_K$ is the kinetic temperature, $\mu_{NH_3} = 17.03$ is the mean molecular weight of \ammonia, $m_H$ is the atomic hydrogen mass, and $\sigma_{obs}$ is the total \emph{corrected} velocity dispersion of the lines, $\sigma_{obs} = \Delta{v_c}/\sqrt{8\ln{2}}$ for $\Delta{v_c}^2 = \Delta{v}^2 - v_{ch}^2$. Using the \ammonia\ study by \citet{Rosolowsky08}, we adopted $T_K = 11 \mbox{ K}$ for all our targets except B1-E2, where we used our derived gas temperature, $T_K = 10.27 \mbox{ K} \pm 0.38$ K.

\subsubsection{CCS and \HCFN}

\begin{figure*}
\centering
\includegraphics[scale=0.9]{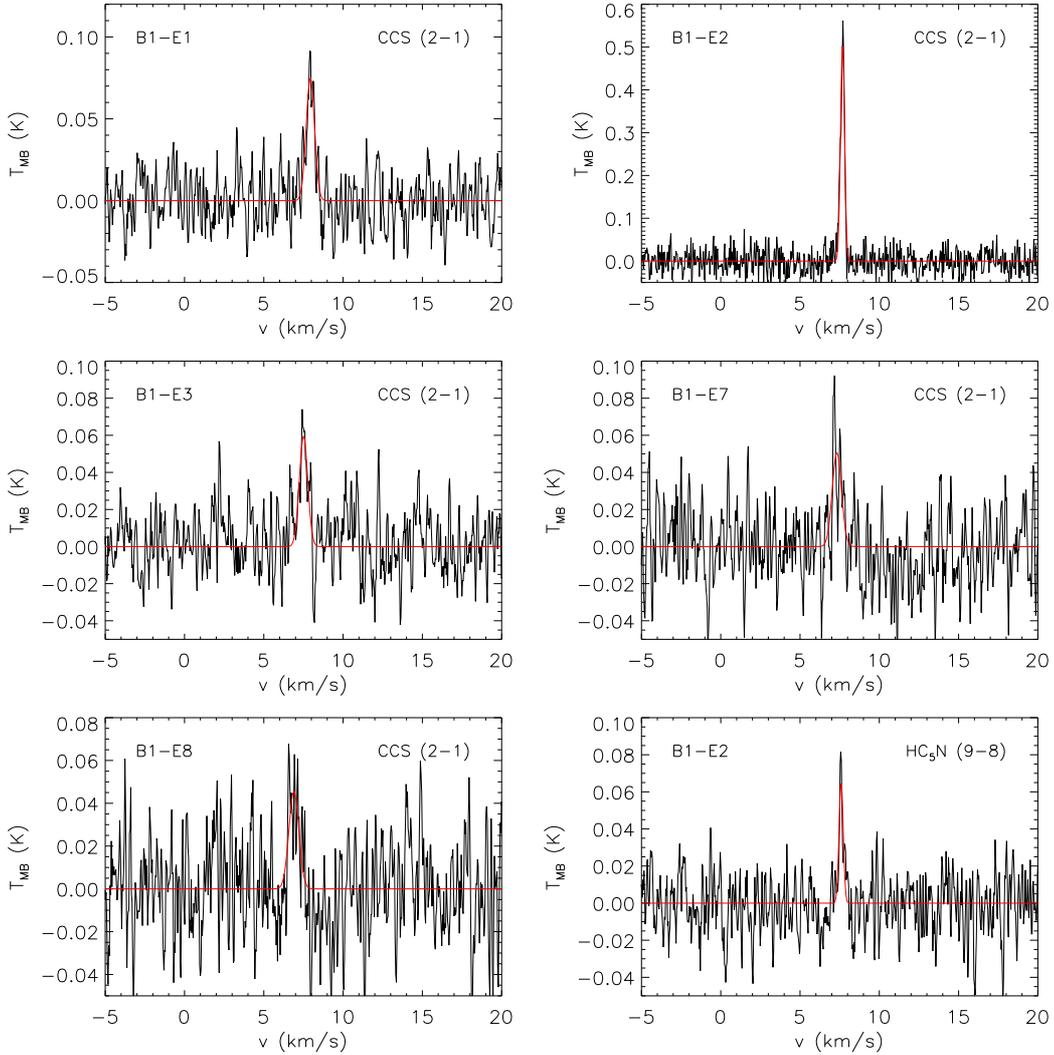}
\caption{Fits to the detected \ccs\ and \HCFN\ (9-8) spectra. The weaker \ccs\ spectra and the B1-E2 \HCFN\ (9-8) spectra were all smoothed by a boxcar with a 2-channel width to improve the visual appearance of the lines. The B1-E2 \ccs\ spectrum was not smoothed. \label{ccs_hc5n}}
\end{figure*}

\ccs\ was detected towards five targets (see Table \ref{pointings}) and  \HCFN\ (9-8) was detected towards only B1-E2. Figure \ref{ccs_hc5n} shows examples of these spectra.  We fit the \ccs\ and \HCFN\ (9-8) spectra with single Gaussians. The velocities determined for most \ccs\ detections and the single \HCFN\ (9-8) detection were similar to those found from \ammonia, suggesting these dense gas tracers are probing the same region as the \ammonia\ emission. The \ccs\ spectra for B1-E1 and B1-E8, however, revealed fairly different values. For B1-E1, the \ccs\ line emission appears redshifted by $\sim$ 0.3 \kms\ (14 $\sigma$) with respect to the \ammonia\ (1,1) line emission, whereas for B1-E8, the \ccs\ line emission is redshifted by $\sim 0.1$ \kms\ (1.5 $\sigma$). The redshifted \ccs\ emission in B1-E1 may be tracing a different region of dense gas from the \ammonia\ emission, such as a region undergoing infall \citep[e.g., see][]{Swift05}.
3

\section{Discussion}\label{discussion}

Our \emph{Herschel} data have revealed extensive substructures within B1-E for the first time. Such faint substructures may represent an early evolutionary stage in star formation, where a first generation of cores is forming from lower density material. Thus, it is important to characterize the substructures seen in these \emph{Herschel} maps. 

\subsection{Comparison with Jeans Instability}\label{jeans}

The minimum length scale for gravitational fragmentation in a purely thermal clump or cloud is described by the Jeans length, $\lambda_J = (c_s^2\pi/G\rho)^{1/2}$, where $c_s$ is the thermal sound speed, $G$ is the gravitational constant, and $\rho$ is the density of material \citep{StahlerPalla05}. In terms of temperature and number density, we find,
\begin{equation}
\lambda_J = 0.21 \left(\frac{T}{10\ \mbox{K}}\right)^{0.5}\left(\frac{n}{10^4\ \vol}\right)^{-0.5} \mbox{pc}.
\end{equation}
Ideally, the Jeans length should be calculated for the gas temperature and gas density. In small, dense scales \citep[i.e., $\gtrsim 10^5$ \vol;][]{difran07}, the gas and dust temperatures are coupled, whereas in larger, more diffuse scales, the gas and dust are decoupled and the gas may be warmer than the dust \citep{young04,difran07, Ceccarelli07}. The central region of B1-E containing the substructures, i.e., the area contained within $\NHH > 6 \times 10^{21}$ \cden, has a mass of $\sim 110$ \Msun\ and an effective radius ($R_{eff} = \sqrt{A/\pi}$) of $\sim 0.46$ pc.  Approximating B1-E as a sphere, we find an average density of $\sim 5 \times 10^3$ \vol, suggesting that the the gas and dust in B1-E are likely decoupled such that the mean dust temperature of $\sim 14$ K (see Section \ref{seds}) is a lower limit to the gas temperature. Assuming a lower limit gas temperature of 14 K and $n = 5 \times 10^3$ \vol, the lower limit $\lambda_J \approx 0.35$ pc. Note that the kinetic temperatures measured in Section \ref{ammoniaSect} correspond to the smaller, denser scales where the gas and dust temperatures should be coupled and not the scales measured here.

The minimum projected separations between our nine substructures are generally between 0.1 pc and 0.19 pc, with a median value of $\sim 0.13$ pc for all nine sources. Recall, however, that B1-E6 and B1-E7 have an angular separation of $\sim 0.05$ pc, a factor of 2 closer than any other pair of substructures. Despite their close proximity, B1-E6 and B1-E7 show different centroid velocities (see Table \ref{ammonia}), and therefore, cannot be considered a single object. Also, we can only measure the projected 2-D separations between our nine substructures and not their physical 3-D separations. Assuming a typical inclination angle of 60$\degree$, our median separation is $\sim 0.15$ pc for all nine substructures or $\sim 0.17$ pc excluding B1-E6 and B1-E7.

The median minimum separation is a factor of two smaller than our best estimate of the Jeans length, though our Jeans length estimate is, itself, uncertain within a factor of a few. The fact that we find a minimum separation that is less than the Jeans length is still interesting and could indicate gravitational fragmentation followed by bulk contraction of the B1-E group, but the overall uncertainties in determining the minimum separations and Jeans length make this difficult to validate. Additionally, B1-E does not appear to be filamentary (see Figure \ref{targets}) and therefore, the clump could be extended along the line of sight resulting in more significant radial distances between substructures. 

For the substructures, we use the Jeans mass to explore their thermal stabilities. Similar to the Jeans length, the Jeans mass is a critical mass scale for gravitational fragmentation \citep{StahlerPalla05}. This critical mass can be described as,
\begin{equation}
M_J = 2.9 \left(\frac{T}{10\ \mbox{K}}\right)^{1.5}\left(\frac{n}{10^4\ \vol}\right)^{-0.5} \Msun.
\end{equation}
The B1-E substructures have average densities of $n \sim 1 \times 10^5$ \vol\ (see Section \ref{substructures}). Furthermore, we found a gas temperature for B1-E2 of $\sim 10$ K. Note, that this gas temperature corresponds to the denser B1-E2 substructure and should not be assumed for the lower density B1-E clump. Assuming $T = 10$ K and $n = 1 \times 10^5$ \vol, the critical Jeans mass is $M_J \approx 0.9$ \Msun. Several of the substructures have mass limits on order of the Jeans mass (see Table \ref{properties}), suggesting that these sources are approaching a critical, unstable mass. Thus, the B1-E substructures are interesting prospects for future studies of substructures evolution.

\subsection{Time Scale for Interactions}\label{timescale}

From the \emph{Herschel} observations, the B1-E substructures appear linked (see Figure \ref{targets}), which might indicate that they are interacting.  The timescale for any two substructures to interact can be estimated from the typical separation between the substructures and their velocity dispersion. For simplicity, we adopt the average projected separation between all nine substructures of  $\sim 0.40$ pc. For an inclination angle of 60$\degree$, the typical separation between the substructures is $d_{ave} \sim 0.46$ pc. To estimate the velocity dispersion of the sources, we use the dispersion of the centroid velocities of \ammonia\ (1,1) from all nine objects. The weighted 1D centroid velocity dispersion is $\sigma_{1D} \approx 0.24\ \kms$ and assuming symmetry, the 3D velocity dispersion\footnote{Assuming $\sigma_{3D}^2 = \sigma_{r}^2 + \sigma_{\phi}^2 + \sigma_{\theta}^2 = 3\sigma_r^2$.} is $\sigma_{3D} \approx 0.42\ \kms$. Thus, the B1-E substructures have an interaction timescale of $\sim 1$ Myr, a factor of two larger than recent estimated prestellar core lifetimes and less than the expected protostellar core lifetime \citep[e.g.,][]{JKirk05, Ward-T07, Enoch08}. Furthermore, this timescale corresponds to the interaction time between any two substructures and not the interaction time between nearest neighbour substructures. Given that the nearest neighbour separation is a factor of $\sim$ 3 smaller than the typical separation adopted here, these substructures have the potential to interact prior to forming stars. Thus, competitive accretion may be significant during further evolution of B1-E \citep[e.g.,][]{Bonnell01, Krumholz05, Bonnell06}.

\subsection{Comparison with Virial Equlibrium}

Not all dense substructures will form cores and stars. For example, a core with kinetic energy from internal motions that exceeds its gravitational binding energy may be transient and unable to form persistent dense structures or stars. According to \citet{Mckee99}, a core is gravitationally bound if the virial parameter, $\alpha$, is
\begin{equation}\label{virialEq}
\alpha = \frac{M_{virial}}{M} = \frac{5\sigma_{v}^2R}{GM} \lesssim 2
\end{equation}
where $\sigma_{v}$ is the velocity dispersion, $R$ is the radius, and $M$ is the mass.

B1-E has an effective radius of 0.46 pc (see Section \ref{jeans}). We estimate the velocity dispersion using the centroid velocity dispersion from Section \ref{timescale}, $\sigma_{3D} = 0.42$ \kms. This velocity dispersion was derived from the motions of the individual substructures and thus, reflects the turbulence in the cloud that first created the substructures rather than the thermal velocity dispersion of the gas. The expected thermal velocity of the gas is less than 0.42 \kms, and thus non-thermal support is important. Assuming $R = 0.46$ pc and $\sigma_v = 0.42$ \kms, B1-E has a virial mass of $\sim 96$ \Msun, or $\alpha \approx 0.9$ for a clump mass of 110 \Msun\ (see Section \ref{jeans}). These results imply that B1-E is itself gravitationally bound and thus, likely to form dense cores and stars in the future.

For the individual substructures, velocity dispersions were determined from \ammonia\ (1,1) line widths, after estimating the non-thermal component ($\sigma_{NT}$; see Table \ref{ammonia}) and the thermal component ($c_s$) for a mean molecular weight of $\mu_{H_2} = 2.33$. Table \ref{virial} shows our values for the total velocity dispersion, $\sigma_{v}$, the ratio of the velocity dispersion to the thermal sound speed, the virial mass,$M_{virial}$, and the virial parameter, $\alpha$, for each substructure. Most of the virial parameters are $\gg 2$, suggesting that these substructures themselves are not gravitationally bound and thus, may not necessarily represent persistent objects. B1-E2 has the smallest virial parameter, $1 \lesssim \alpha \lesssim 3$, however, and may be itself gravitationally bound. B1-E8 and B1-E3 have lower limit estimates of $\alpha \sim 3$ and given that we know $\alpha$ only within a factor of a few, these sources may also be gravitationally bound. These virial limits, however, depend greatly on the source mass, which can vary by a factor of $\sim 4$ depending on our treatment of the LOS material (see Section \ref{substructures}).  To be bound, these candidates must have negligible LOS material towards them, which is very unlikely. Thus, B1-E2 is the only strong candidate for being gravitationally bound.

\begin{table}
\caption{Virial Analysis Properties}
\label{virial}
\centering
\begin{tabular}{lcccc}
\hline\hline
Source &  $\sigma_{v}$ & $\sigma_{v}/c_s$\tablefootmark{a} & $M_{virial}$ & $\alpha$\tablefootmark{b} \\
              & (\kms) &  &  (\Msun)    &\\
\hline
B1-E1 & 0.460 $\pm$ 0.030 & 2.3 $\pm$ 0.3 & 8.5  $\pm$ 4.8 &  5 - 14\\
B1-E2 & 0.216 $\pm$ 0.005 & 1.1 $\pm$ 0.03 & 1.8 $\pm$ 0.9 &  1 - 3\\
B1-E3 & 0.342 $\pm$ 0.022 & 1.7 $\pm$ 0.2 & 5.1 $\pm$ 2.9  &  3 - 10\\
B1-E4 & 0.493 $\pm$ 0.024 & 2.5 $\pm$ 0.3 & 8.2 $\pm$ 4.5  & 8 - 27 \\
B1-E5 & 0.397 $\pm$ 0.018 & 2.0 $\pm$ 0.2 & 8.3 $\pm$ 4.5  & 4 - 17\\
B1-E6 & 0.424 $\pm$ 0.022 & 2.2 $\pm$ 0.2 & 6.0 $\pm$ 3.3 &  6 - 20 \\
B1-E7 & 0.351 $\pm$ 0.023 & 1.8 $\pm$ 0.2 & 4.1 $\pm$ 2.3  & 5 - 14\\
B1-E8 & 0.291 $\pm$ 0.021 & 1.5 $\pm$ 0.2 & 3.4 $\pm$ 2.0  & 3 - 11\\
B1-E9 & 0.343 $\pm$ 0.025 & 1.7 $\pm$ 0.2 & 3.6 $\pm$ 2.1  & 4 - 12\\
\hline
\end{tabular}
\tablefoot{
\tablefoottext{a}{Velocity dispersion divided by sound speed, assuming $T = 11$ K for all substructures except B1-E2 where $T = 10.27$ K.}
\tablefoottext{b}{Virial parameter considering the upper and lower mass limits for each substructure (see Table \ref{properties}).}
}
\end{table}

 Most B1-E substructures have very large virial parameters and are thus, expected to be unbound. In comparison, the Jeans mass (see Section \ref{jeans}) is a mass scale for only gravitational fragmentation in a thermal clump and does not include the contributions from non-thermal support. Several of the substructures with $\alpha \gg 2$ have masses on order of the Jeans mass, $\sim 0.9$ \Msun, suggesting that the Jeans mass analysis is too simplistic and does not well represent the evolutionary state of these structures. For example, the substructures may have significant turbulent motions. With only one potentially bound substructure (B1-E2), it is difficult to determine if B1-E will eventually form several cores or no cores at all (i.e., these substructures are all transient).

The substructures themselves show large line widths (see Table \ref{virial}), atypical of dense prestellar cores, which are subsonic. In particular, B1-E2 has the narrowest line profiles (by at least a factor of 2) and the strongest line detections, suggesting that this object is the most evolved of the substructures. These observations hint at a possible dynamic evolution, where dense cores first contain significant non-thermal, turbulent motions which dissipate into coherent, quiescent dense cores. If so, the substructures in B1-E may become more bound over time as their turbulent support is dissipated, a process which may explain the narrower lines toward B1-E2.

Note that we assume $T_K = 11$ K for our substructures (except B1-E2). In Section \ref{ammoniaSect}, we found upper limits for kinetic temperature $> 11$ K (see Table \ref{ammonia}). Furthermore, we are using a simplified virial equation in Equation (\ref{virialEq}). Ideally, the virial equation includes magnetic pressure and surface pressure terms which can either assist gravity in the collapse or help in the support of a clump \citep[e.g., see][]{NakamuraLi08}. Our observations, however, do not measure these quantities and thus, we use a simplified virial equation. Nevertheless, B1-E2 is special with respect to the other substructures. Indeed, we may have the first observations of a core precursor. 

\subsection{Comparison with Other Star Forming Regions}\label{comp_obs}

B1-E has column densities well above $\sim 5 \times 10^{21}$ \cden, the core formation threshold reported by \citet{Kirk06}. Since the other clumps with similarly high column densities are actively forming stars, it is reasonable to expect B1-E to do the same in the near future. Furthermore, \citet{Lada10} found a good correlation between the number of YSOs in a cloud (or clump) with the cloud mass above $A_K > 0.8$ (or $A_V > 7.3$) for several star forming regions. Based on their extinction threshold, we find a clump mass of  $\sim 90$ \Msun\ for $\NHH > 7.3 \times 10^{21}$ \cden\ and thus, we could expect $\sim 10$ YSOs to form in B1-E. With \emph{Herschel}, we have detected several substructures with moderate densities of $\sim 10^5$ \vol\ (see Table \ref{properties}), similar to what is often defined for dense cores, i.e., $\gtrsim 10^4$ \vol\ \citep{BerginTafalla07}.

These substructures, however, were not well observed by SCUBA at 850 \um\ or Bolocam at 1 mm (see Figure \ref{b1e850_250}), and thus, are not dense cores in the traditional definition. The substructures in B1-E also have supersonic velocity dispersions. In general, low-mass starless cores have subsonic velocity dispersions \citep[e.g.,][]{Myers83, Kirk07, Andre07, Pineda10} whereas protostellar cores can have supersonic velocity dispersions \citep[e.g.,][]{Gregersen97, difran01}. Since we see no evidence of protostellar activity whatsoever (i.e., from \emph{Spitzer} data), these substructures are likely unevolved.

Thus, the substructures in B1-E are likely core precursors and more observations of this region should provide some insight into core formation (transient or persistent) in molecular clouds. In particular, B1-E is isolated from YSOs and young stars, so the region is unaffected by environmental processes such as outflows or winds. A nearby expanding dust shell, seen in IRAS 60 \um\ and 100 \um\ data \citep{Ridge06dustRing}, may be sweeping material towards B1-E, though the ring itself does not appear to be directly interacting with this clump. This shell may have a larger impact on the evolution of structure in B1-E in the future.

A relatively pristine core forming region is exceedingly rare. The only additional case may be the inactive high extinction clump (211) identified in L1495 by \citet{Schmalzl10}, though further detailed studies of that clump are still necessary to determine what, if any, substructures are forming cores and stars. A collection of starless cores in Aquila form a similar grouping as in B1-E, however this region has two known protostars nearby ($\lesssim$ 1 pc) and so is not as pristine as B1-E \citep[][see their Figure 5]{Konyves10}.

\subsection{Comparison with Core Formation Models}\label{comp_theo}

Core formation simulations attempt to recreate molecular cloud fragmentation, implementing factors such as gravitational collapse, magnetic pressure, and turbulent energy. Many recent studies have argued that models of core formation must consider both ambipolar diffusion and turbulence \citep[e.g.,][]{LiNakamura04, Dib07, NakamuraLi08, Basu09, Basu09b, GongOstriker11}. Magnetic fields introduce an additional pressure support that opposes gravitational collapse, whereas turbulence either opposes collapse by introducing kinetic energy support or induces collapse via energy dissipation from compressive shocks.

Several recent studies have examined the effects of different magnetic field and turbulent flow properties \citep[e.g.,][]{LiNakamura04, Basu09, PriceBate09}. In particular, these simulations examine fragmentation in subcritical (where magnetic fields dominate over gravity) and supercritical (where gravity dominates over magnetic fields) regimes with dissipating turbulent flows. In the \emph{subcritical} simulations, the fragmenting clouds generally form cores in relative isolation and with less filamentary morphologies, such as ring-like distributions of loose core groups \citep[see also][]{Li02}. These cores tend to form more slowly, i.e., fragmentation is suppressed by the strong magnetic field, and since large-scale turbulent flows are generally damped, the velocity field is generally subsonic. \citep[Note that][find that strong magnetic fields can induce oscillations and create a supersonic velocity field, i.e., $\lesssim 3c_s$.]{Basu09} In contrast, \emph{supercritical} simulations generally form cores associated with filaments, and these cores tend to form relatively quickly and can move supersonically through their environment.

An isolated core forming region is necessary to test model predictions and constrain initial conditions of dense gas before the onset and influence of nearby stars and internal protostars. Thus, B1-E is an ideal target for such measurements (see Section \ref{comp_obs}). From our observations, B1-E contains a small, loose grouping of substructures. Although the substructure morphology observed in B1-E is not entirely ring-shaped, the distribution hints at an inclined ring-like structure. Furthermore, if star formation is triggered, it is surprising that B1-E, which is bookended by two highly active and more evolved clumps, IC 348 to the far east and NGC 1333 to the far west, contains no evidence for evolved star formation itself. Furthermore, IC 348 shows many evolved pre-main sequence stars and NGC 1333, though appearing younger, has many YSOs \citep{Bally08, Herbst08, Walawender08}, suggesting that B1-E is not part of an age gradient.

These observations suggest that \emph{B1-E is influenced by a strong, localized magnetic field.} Further tests of this scenario would be possible by observing the entire velocity field for B1-E. Instead, we could only estimate the bulk velocity field from the relative motions of the individual substructures (from line centroids; see Section \ref{timescale}), $\sigma_{3D} = 0.42$ \kms. Additionally, we do not have direct measurements of the gas temperature for the B1-E clump. If we use the mean B1-E dust temperature of 14 K as the lower limit gas temperature, $c_s \gtrsim 0.22$ \kms\ and the velocity field within B1-E may be supersonic by $\lesssim 2c_s$. These motions can arise in either subcritical or supercritical clouds, though the latter is generally associated with supersonic velocities.

Direct measurements of the magnetic field would also be useful to constrain its role in B1-E. Unfortunately, very little magnetic field information is available here. \citet{Goodman90} measured optical polarization towards all of Perseus, and for B1-E, they found strong, ordered polarization vectors reaching $\sim 9$\%\ polarized light, a factor of two higher than any other dense region in Perseus. While these polarization observations may suggest a strong magnetic field is associated with the B1-E region, these observations are dependent on the field viewing angle. Also, these data were obtained at very low resolution resulting in only three vectors coinciding with B1-E. 

A strong magnetic field does not necessarily relegate turbulent compression or gravity to minor roles in the evolution of B1-E. Indeed, \citet{NakamuraLi08} suggest that core forming regions exhibit different phases where magnetic fields, gravity, and turbulence are sometimes dominant and sometimes secondary. For example, they suggest that core forming regions begin with strong magnetic fields and strong turbulence, such that gravity is a secondary effect. After the turbulent energy dissipates, gravitational collapse can proceed via magnetic field lines.

\section{Conclusions}\label{conc}

With recent \emph{Herschel} dust continuum observations from the \emph{Herschel} Gould Belt Survey, we identified substructure in the Perseus B1-E region for the first time. With our \emph{Herschel} data, we determined the temperature and column density across the region. We selected the nine highest column density substructures for complementary observations with the GBT. We summarize our main conclusions as follows:

\begin{enumerate}

\item B1-E contains a loose collection of roughly a dozen prominent substructures. This morphology is atypical of most star forming regions, which produce dense clusters and organize material along filaments. Such a loose collection of cores can be produced in magnetically subcritical simulations \citep[e.g., see][]{Li02, Basu09} influenced by a strong magnetic field. Furthermore, a strong magnetic field will delay the onset of star formation, which may explain the age discrepancy between B1-E and the other nearby clumps in Perseus, IC 348 and NGC 1333.

\item The B1-E clump as a whole is gravitationally bound with an estimated virial parameter of $\alpha \approx 0.9$. Therefore, B1-E is likely to form dense cores and stars. Assuming $T = 14$ K and $n = 5 \times 10^3$ \vol, we find a Jeans length within the entire B1-E clump of $\sim 0.35$ pc, accurate within a factor of a few. We measure a median nearest neighbour separation of $\sim 0.15$ pc for our nine substructures ($\sim 0.17$ pc excluding B1-E6 and B1-E7), for an inclination of 60$\degree$. Thus, the substructures have a median minimum separation that is less than the Jeans length by a factor of two. This smaller length scale could indicate that B1-E contracted after the observed substructures formed.

\item Several of the B1-E substructures have masses on order of the Jeans mass ($\sim 0.9$ \Msun), assuming $T = 10$ K and $n = 1 \times 10^5$ \vol. Nevertheless, most substructures have large virial parameters ($\alpha >> 2$) indicating that they are gravitationally unbound. The large virial parameter suggests that non-thermal motions are significant and that the Jeans mass does not well represent the critical mass scale for these substructures. B1-E2 is the only substructure with a small virial parameter ($1 \lesssim \alpha \lesssim 3$) and, due to its narrow line emission, may be gravitationally bound. Furthermore, \ammonia\ (2,2) and \HCFN\ (9-8) were only detected towards B1-E2. Thus, B1-E2 is an excellent candidate for a core precursor. 

\item The B1-E substructures have a substantial centroid velocity dispersion ($\sim 0.42$ \kms) resulting in an interaction timescale of $\sim 1$ Myr, assuming an average separation of $\sim 0.46$ pc. This timescale is well within the lifetime of protostellar cores. Thus, competitive accretion may play a significant role in the evolution of structure in B1-E.

\end{enumerate}

The B1-E region appears to be an excellent candidate for future core formation. Further studies, however, are necessary to explore the dust and dynamics of the bulk gas in B1-E. For example, additional submillimetre continuum data along the Rayleigh-Jeans tail (e.g., with SCUBA-2) will improve SED-fitting and more accurately probe the dust emissivity index, $\beta$, size, and mass. To date, our \emph{Herschel} observations of B1-E are the only well detected continuum data for the region. Additionally, our GBT observations are the only high resolution kinematic data for B1-E, and we observed only the nine highest column density substructures seen with \emph{Herschel}. These data illustrate the potential of B1-E to be a core forming region. Future observations of \emph{both} the magnetic field strength and the turbulent velocities are necessary to probe further how B1-E will evolve.

\vspace{1cm}
\begin{acknowledgements}
We thank the anonymous referee for comments that greatly improved the discussion and narrative of this paper. This work was possible with funding from the Natural Sciences and Engineering Research Council of Canada CGS award. We thank B. Ali, S. Basu, R. Friesen, D. Johnstone, V. Konyves, G. Langston, A. Pon, S. Schnee, B. Schulz, N. Wityk, and the kind folks at the NHSC helpdesk and at the GBT for their invaluable assistance and advice throughout this project. We also thank the CITA IT staff for their time and attention. SPIRE has been developed by a consortium of institutes led by Cardiff Univ. (UK) with Univ. Lethbridge (Canada); NAOC (China); CEA, LAM (France); IFSI, Univ. Padua (Italy); IAC (Spain); Stockholm Observatory (Sweden); Imperial College London, RAL, UCL-MSSL, UKATC, Univ. Sussex (UK); Caltech, JPL, NHSC, Univ. Colorado (USA). This development has been supported by national funding agencies: CSA (Canada); NAOC (China); CEA, CNES, CNRS (France); ASI (Italy); MCINN (Spain); SNSB (Sweden); STFC (UK); and NASA (USA). PACS has been developed by a consortium of institutes led by MPE (Germany) with UVIE (Austria); KU Leuven, CSL, IMEC (Belgium); CEA, LAM (France); MPIA (Germany); INAF-IFSI/OAA/OAP/OAT, LENS, SISSA (Italy); IAC (Spain). This development has been supported by the funding agencies BMVIT (Austria), ESA-PRODEX (Belgium), CEA/CNES (France), DLR (Germany), ASI/INAF (Italy), and CICYT/MCYT (Spain). The Green Bank Telescope is operated by the National Radio Astronomy Observatory. The National Radio Astronomy Observatory is a facility of the National Science Foundation, operated under the cooperative agreement by Associated Universities, Inc.
\end{acknowledgements}

\bibliographystyle{aa}
\bibliography{references}

\end{document}